# A New Approach to Improve the Performance of OFDM Signal for 6G Communication

Usha S.M[1] and Mahesh H.B[2]

[1]Department of Electronics & Communication Engineering, JSS Academy of Technical Education, Bengaluru, Visveswaraya Technological University, India
[2]Department of Computer Science Engineering, PES University, Bangalore, India

*ABSTRACT*

*The orthogonal frequency division multiplexing is a very efficient modulation technique that can achieve very high throughput by transmitting many carriers simultaneously and it is spectrally efficient because of the proximity of the subcarriers. OFDM is used in 5G communication for higher data transmission. 6$^{th}$ generation communication also demands OFDM, since it is more spectrally efficient and suitable for high data transmission. The drawback of the OFDM includes peak to average power ratio and sensitivity to carrier offsets and drifts. The usage of a non-linear power amplifier causes the signal spreading and leads to inter-modulation and signal constellation distortion. These two distortions have an impact on the signal-to-noise ratio and hence reduce the efficiency. The methods used to reduce PAPR are clipping and filtering, selective mapping, partial transmit sequence, tone reservation, and injection and non-linear commanding. The drawbacks of the above methods are computational complexity, spectrum inefficiency, increase in bit error rate and PAPR rate. In this work, three effective methods are discussed and compared to improve the performance parameters. These are adaptive peak window method based on harmonize clipping, harmonics kernel adaptive filter and Slepian-based flat-top window techniques are presented to reduce the BER, PAPR, and CCDF to improve the signal-to-noise of the system. This window technique averages out the noise spread out in the spectrum and thus reduces the signal loss by minimizing peak to average power ratio. The results are analyzed and compared with the existing conventional methods. Finally, the reductions in PAPR, BER and CCDF obtained are discussed in the results and comparison section. The proposed work has a higher signal-to-noise ratio than the conventional methods.*

## 1. INTRODUCTION

Bandwidth can be efficiently utilized by transmitting the signals orthogonally. The signals transmitted from the earth stations have to propagate for longer distances to reach the satellite in space. When the signal is propagating, it is affected by many environmental parameters. The signal collides with the taller buildings, hill stations, long towers, etc. It decreases signal strength due to fading and inter-symbol interference. The fading can be minimized by sending signals orthogonally but this increases the peak to average power ratio [1-3]. PAPR increases the interference level and decreases the system performance [4-5]. Orthogonal multiplexing is a multi-carrier modulation technique. It is more spectrally efficient, robustness against multi-path signal fading and supports large data transmission. As a result, OFDM is widely used in many radio communication applications. OFDM transmits the data simultaneously by splitting large data streams into smaller data streams via many subcarriers [6-7]. OFDM also minimizes the dispersion that occurs due to the signal spreading. The reason for signal spreading is the inter-symbol interference and cross talk. The signal spreading due-to time delay is minimized by sending signals orthogonally [8-9].

The implementation of OFDM is shown in Figure 1. At the transmitter, the samples X0, X1, X2, X3…….XN-1 applied to IDFT block, the IDFT converted signal is passed through a channel. At the receiver, DFT is performed on incoming IDFT data samples. The composite OFDM data block is defined by Eq. (1).

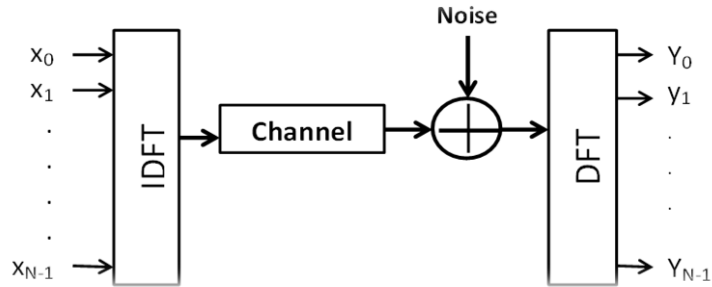

Figure 1. OFDM Implementation

$$x(t) = \frac{1}{\sqrt{N}} \sum_{n=0}^{N-1} X_N \, e^{-j2\pi n \Delta f t} \quad 0 \le t \le NT \tag{1}$$

Where j = -1 represents the useful data block period. f is the separation stuck between the subcarriers, and NT symbolizes the duration of the desired information chunk. PAPR is the ratio of maximum signal power to the average energy of the signal power. If the maximum signal power max $\{|xn|2\}$ is reduced. The PAPR can be reduced correspondingly. The maximum power is directly proportional to the PAPR as shown in Eq. (2).

$$\text{PAPR} = \text{MAX } [X(N)]^2 / \, E[|X(N)|^2] \tag{2}$$

The power amplifiers at the transmitting end are driven towards aturation; this reduces the signal-to-noise ratio and results in the reduction of the BER performance. As a result, it is better to handle excessive PAPR by decreasing signal peak power. Different methods for PAPR reduction are represented in this paper. The performance of the PAPR reduction is evaluated through the complementary cumulative distribution and bit error rate performance.

## 1.1. Clipping and Filtering

Clipping the orthogonal frequency division multiplexing signal's peak before running it to the power amplifier is one of the simplest PAPR reduction strategies as shown in Figure 2. Clipping is a nonlinear phenomenon that causes distortion. Intra-modal interference is reduced by filtering the signal after clipping. An oversampling of the signal by using IFFT reduces the intra-modal dispersion by transmitting part of the noisy signal outer the information band. The bit error rate performance is improved by filtering the OFDM signal and preserving the efficiency of the spectrum by increasing the peak power.

An input data stream is modulated using quadrature phase-shift keying, on the modulated data the orthogonal frequency division multiplexing is employed. The OFDM modulated data is passed through a clipping circuit and filtered using an FIR filter. The filtered data is demodulated and information bits are retrieved. The noise results during the clipping process cause intra-band distortion. This distortion cannot be minimized using filtering, so oversampling is carried out to minimize the intra-band distortion.

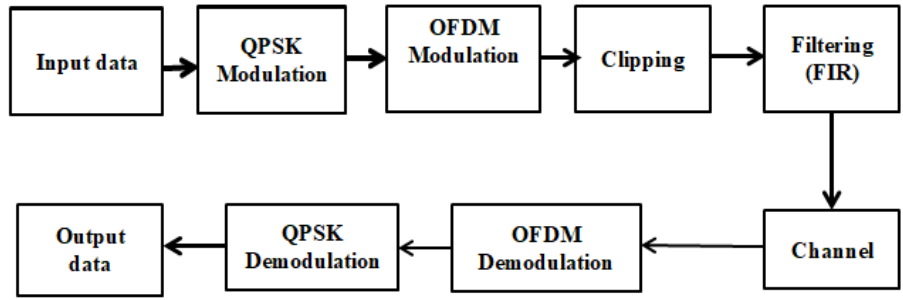

Figure 2. OFDM with clipping and filtering

The input data stream is modulated using QPSK and multiplexed using OFDM. The signal is clipped and filtered using FIR filter. At the receiver, data is demodulated and information bits are retrieved. Then, at amplitude A, the samples, x, are clipped as follows as in Eq. (3), (4) and (5).

$y = -A$, if $x < -A$     (3)
$y = x$, if $-A \leq x < A$     (4)
$y = A$, if $x < A$     (5)

### 1.2. Selective Mapping

The selective mapping is shown in Figure 3. In this method, binary data is mapped using the QPSK mapping method. The mapping data sequence is multiplied with the phase data. The output of the multiplier is applied to the IDFT block.

The data is compared and one with low PAPR is selected. At the receiver, the operation is performed to retrieve the data. The OFDM symbols can be brought forth by multiplying the un-derived data block by M diverse phase sequence $P_m$, each of length N.

The IDFT of the component is as shown in Eq. (6). The data having lowest PAPR is selected.

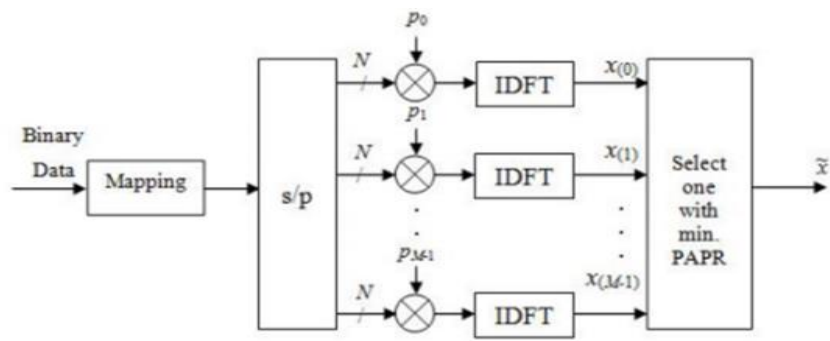

Figure 3. OFDM transmitter block diagram with SLM

$$x_m = IDFT[X_1 e^{j\varphi_{m,1}} \; X_2 e^{j\varphi_{m,2}} \ldots X_N e^{j\varphi_{m,N}}] \qquad (6)$$

### 1.3. Partial Transmit Sequence

In this method, block of data is divided into various non-overlapping data sub-blocks and each sub-block is carried out separately. Each sub-block is multiplied with the phase weighting factor and the optimizer block selects the one with a low paper level as in Figure 4.

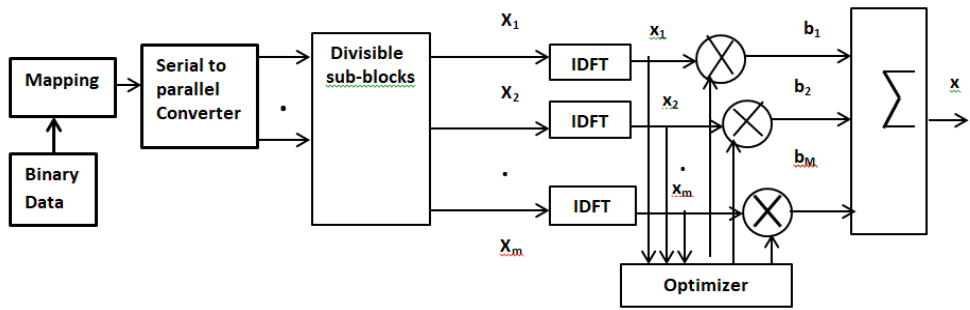

Figure 4. The partial transmit sequence method

The paper is organized into six different sections. In the first section of this paper, an introduction to the work and various existing models to reduce the PAPR is discussed. In section II, related work is represented. In section III, the proposed work is discussed. The results obtained are discussed in section IV. The results comparison and analysis are provided in section V and conclusions are stated in section VI.

## 2. RELATED WORK

This paper explored various methodologies to reduce the peak to average and bit error rate ratio [10]. The author mainly focused on the partial transmits sequence technique to improve the performance by reducing the peak to average power ratio and error value. The work was carried out using MATLAB. This work states fractional improvement in PAPR and BER rate.

The work was carried out with multiple inputs and multiple output-orthogonal frequency division multiplexing systems [11-14]. In this system, a cyclic prefix is encoded with the original data. This method is efficient, if the received signal is exactly matched with the transmitted signal. The decoder circuit is designed specially to reduce the peak to average power ratio. The improvement in performance parameter is noticed but it was not much better than the conventional methods.

This work was carried out to improve the performance of long-term evaluation [15]. The high PAPR in the uplink of the communication channel affects the signal quality. It is an efficient technique to improve performance by mitigating the PAPR at low complexity and with short time delays. This improves the efficiency of the power amplifier. However, the proposed approach did not achieve perfect efficiency, which is a flaw in the work.

In this work [16], PAPR reduction is achieved through ICA algorithm. PAPR reduction can be achieved by accurately designing the ICA algorithm. The time required for calculation and cost factor increases as the subscriber's increases. The required PAPR reduction was not achieved by this method and not suitable for more number of carriers.

In this work, a new method of SPS-SLM technique is designed [17-18]. The data transmission is carried out using the AWGN channel with OFDM. This method solves the problems such as signal spreading, signal cross-talk and system complexity. This works with higher-order quadrature amplitude modulation schemes. This increases data transmission capacity in the wireless communication system. Overall, it improves the data rate efficiency and reduces the complexity of the entire system [19-20]. The disadvantage was that the adapted approach for PAPR reduction is inefficient and the error rate was also high. In [21], the routing strategies to reduce the latency in the 6G network are discussed in this paper. In this paper [22] channel estimation for multiple inputs multiple outputs (MIMO) is presented.

## 3. PROPOSED METHOD

The existing PAPR reduction techniques worked on the principle of a time-domain signal to broadcast from a set of different renderings to lower PAPR, which would reduce performance. There are two types of PAPR mitigation strategies: distortion-free and distortion approach. The PAPR mitigation strategies with distortion are proposed in this work. The three proposed methods are based on clipping and filtering strategies followed by windowing techniques. These are the adaptive peak window method based on harmonize clipping, harmonious kernel adaptive filter, and stochastic similarity flat-top window with Slepian base as discussed in Figure 5. By combining clipping and windowing approaches, the PAPR in OFDM transmission is minimized. Hence bit error rate is also minimized, which solves interference problems.

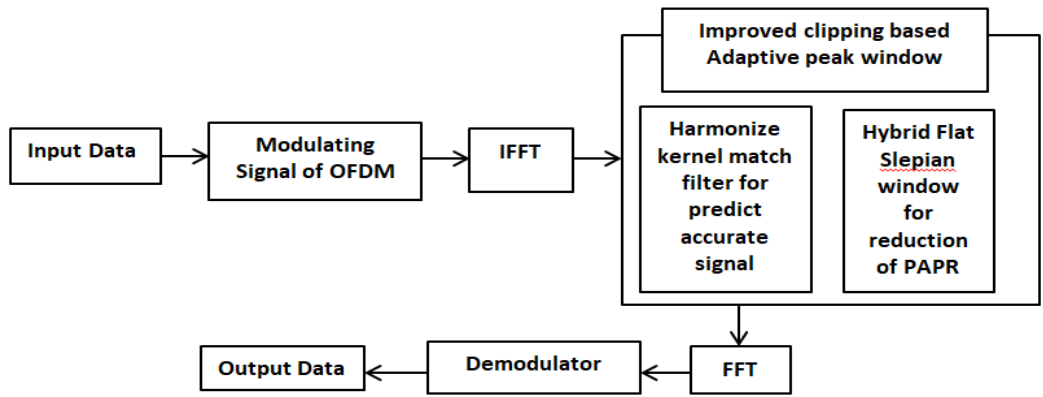

Figure 5. Proposed method block diagram

### 3.1. Adaptive Peak Window Method Based On Harmonize Clipping

A PAPR lowering technology is necessary to reduce the high PAPR at the transmitter to boost power amplifier productivity. As a result, the adaptive peak window with improved clipping has been implemented. A quadrature amplitude modulator modulates the input signal in the OFDM block. This signal is passed through the Inverse Fourier fast transform to convert it from frequency to time domain and then through the clipping/filtering process to maintain the signal's orthogonal environment. This filter keeps the signal orthogonality by lowering PAPR and BER. A matched filter is introduced to predict the phase, which harmonizes the inter-band signals and reduces the distortion. A Slepian-based flat-top window is used to achieve exact noise reduction while maintaining the signals orthogonally. These windows filter out noise in the spectrum and

125

reduce information loss by employing the stochastic resembling technique in precise amplitude measurement at definite intervals of time in discrete form. The signal scattering, PAPR, data loss, computational complexity, and spectrum bandwidth issues are all addressed by this approach making it well-suited for 6G communication systems.

### 3.2. BER reduction with a Harmonious Kernel Adaptive Filter

The harmonious kernel adaptive filter optimizes the signal by maintaining signal orthogonality by reducing PAPR and BER. Kernel prediction provides a reliable signal for detecting conjugated time. To achieve accurate noise reduction while preserving the orthogonality of the signal, a Slepian-based flat-top window is used. By using the stochastic approaches method in exact amplitude measurements at defined intervals of time in discrete form, these windows reduce information loss by filtering out spectral noise. As a result, the above-mentioned OFDM challenges can be overcome. Consider a kernel, which is a two-dimensional symmetric function that is both positive and continuous. This is a function that is both consistently positive. k: X * X -> P. According to "Mercer's theorem", every kernel has a mapping as indicated in Eq. (7).

$$k(x1, x2) = \Phi(x1)\Phi^T(x2) \quad for \ \forall x1, x2 \in X \tag{7}$$

The data is transformed into the feature space using Mercer's theorem and is interpreted as the dot product. This learning system's action can be stated in Eq. (8), (9), (10) and (11) below.

$$\bar{v} = w_n(\bar{x}) = \eta \sum_{i=1}^{n} r_i^a \langle x_i, \bar{x} \rangle X \tag{8}$$

$$r_n^a = v_n - \eta \sum_{i=1}^{n-1} r_i^a \langle x_i, x_n \rangle X \tag{9}$$

$$\Omega_N = \eta \sum_{i=1}^{N} r_i^a \Phi(x_i) \tag{10}$$

The final input-output relationship, after N steps of learning algorithm training, is.

$$\bar{v} = \eta \sum_{i=1}^{N} r_i^a k(x_i, \bar{x}) \tag{11}$$

The kernel filter generates a new kernel element with the input as the midpoint and the coefficient as the difference between the network output and the preferred signal as in Eq. (12). As a result, the signal's orthogonality is maintained and the PAPR is minimized. The kernel adaptive filter's output signal has the same inter band distortion. The matching filter is added to this kernel filter to improve the predictability by minimizing inter-band interference.

$$y(nT) = \begin{cases} \bar{v}(nT) & if \ 0 \leq n \leq N-1 \\ 0 & if \ N \leq n \leq 2N-1 \end{cases} \tag{12}$$

The sample time is represented by the letter T as in Eq.(12). The clipping of the signal in the OFDM system generates interference; signal scattering resulting in signal spreading, which is most common in nonlinear systems. After clipping, the output should be modified with windowing techniques, since clipping action on the signal leads to signal dispersion.

### 3.3. Stochastic Similarity Flat Top Window with Slepian Base

To improve the filtered signal, a Slepian-based flat-top window is used. This approach averages noise in the spectrum by minimizing information loss. The stochastic resemblance technique is used to determine the amplitude accurately at a given interval of time. The flat-top windowing method maintains a virtually flat main lobe and the amplitude deprivation of harmonic components is minimized due to spectral leakage. The side bands are reduced within these windows to prevent interference between distant signal components and/or modest amplitude of the side lobe. The flat-top window receives the adaptive filter's output signal as input. Flat top windowing is the consequence of this as shown in Eq. (13) and (14).

$$s(n) = \sum_{n2=0}^{N} \sum_{n1=0}^{N} w * (n2) w(n1) e^{j2\Pi(n2-n1)t} \tag{13}$$

$$\text{where, } w = \frac{2\Pi n}{N} \tag{14}$$

S(n) values are calculated for different values of n, for n=0, n=1, n=2, n=3, and n=4 respectively. Thus s(0) = 0.215578847, s(1) = 0.416631, s(2) = 0.27263158, s(3) = 0.083578947 and s(4) = 0.006947368. Consequently, the Slepian window is used to improve the signal from the flat top window since its output signal has a condensed frequency resolution and broad noise bandwidth. It specifies the sampling time in seconds, and N denotes the chosen window length in samples. Now that all of the noise has been removed from the waveform, the signal's orthogonality may be carefully maintained by lowering the PAPR. Both clipping and windowing methods are noticed to be successful with OFDM signal. As a result, data loss due to the distortion is decreased. In this work, the kernel filter output is combined with the windowing method to minimize the BER of the signal MinPAPR(x(t)). This strategy of integrating the filtered signal with the windowing technique reduces the noise level by preserving signal characteristics. During the clipping and windowing phase, the time domain signals are ultimately converted to frequency domain signals that are then demodulated using the QAM. The minimum PAPR is calculated as mentioned below in the Eq. (15) and (16).

$$MinPAPR(x(t)) = S(y(nT)) \tag{15}$$

$$w(n) = \sum_{k=0}^{4} (-1)^k a_k \cos(kw) \tag{16}$$

## 4. RESULTS AND DISCUSSION

In the preceding section, the proposed approach is discussed and the implementation results and performance are evaluated. The job is carried out in MATLAB according to the specifications listed below. In a single frame size, the simulation uses 64 subcarriers for a width of 96 bits. Convolution coding is used in this study, and four pilots and a cyclic extension of 25% were used in the 16-QAM modulation technique as in Table 1.

Table 1. Parameter description.

| Parameter Description |
|---|
| The transmission and reception of data involves 64 subcarriers. |
| Size of single frame -96 bits |
| Convolutional Coding is used |
| Scheme for modulation-QAM |
| the total number of pilots-4 |
| Cyclic Extension 25% of the total number of pilots (16) |

### 4.1. PAPR Analysis

A combination of adaptive clipping and windowing techniques is used to reduce PAPR. The following elements are taken under consideration in the analysis: BER, SNR, and the complementary cumulative density function (CCDF). PAPR value of the proposed technique with 8, 12, 32, and 64 subcarriers is respectively 3.72, 4.44, 4.30 and 3.79. This indicates the reduction in the value of PAPR with the proposed method. Figure 6 depicts the graphical representation.

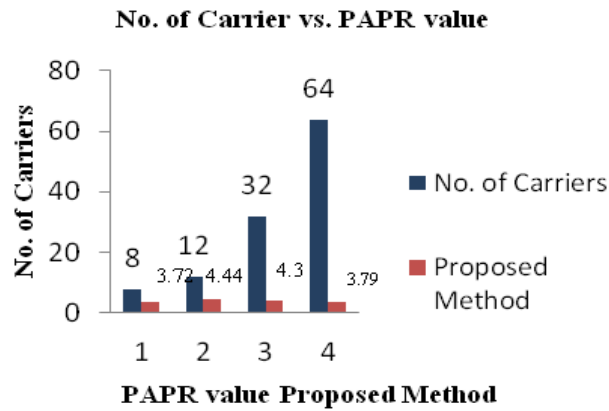

Figure 6. PAPR value against number of carriers

### 4.2. Bit Error Rate Analysis

The bit error rate is the proportion of erroneous bits to the total number of bits transferred from the source terminal to the destination. By lowering the error rate, an efficient system can be created.

Bit error rate = (ST-SR)/ST * 100

Where ST stands for the signal transmission session time, SR stands for the signal receiving time. BER obtained is $415*10^{-3}$, $410*10^{-3}$, $405*10^{-3}$, $375*10^{-3}$ for 8, 12, 32 and for 64 subcarriers respectively. The graphical representation is shown in Figure 7.

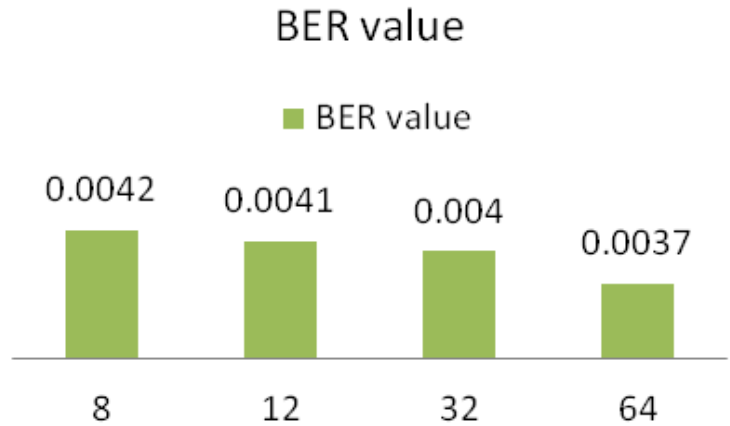

Figure 7. No. of carrier vs. BER value

### 4.3. Signal to Noise Ratio Analysis

The signal-to-noise ratio (SNR) measures the difference between signal and noise power   SNR = 10log10(Psignal/Poise).   The SNR value is 20.09, 21.39, 20.56 and 19.69 for the selected carriers 8, 12, 32 and 64 respectively. The graphical representation is shown in Figure 8.

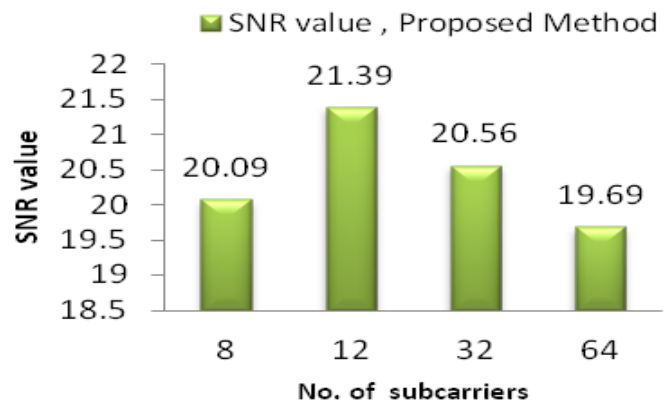

Figure 8. SNR plot of proposed method

### 4.4. Complementary Cumulative Distribution Function (CCDF) Analysis

The CCDF curve shows how much time a signal spends above its average power level, as well as the likelihood that the signal power will be higher than usual. The CCDF value of the proposed method is 4.21, 4.23 4.25 and 4.20 for the carriers 8, 12 32 and 64 respectively. The graphical representation of the CCDF value with carrier count is shown in Figure 9.

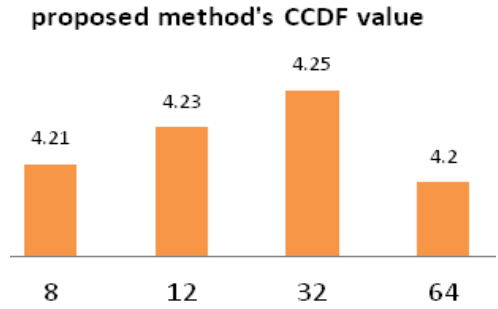

Figure 9. CCDF plot of proposed method

## 5. COMPARISON AND ANALYSIS

The proposed method of PAPR reduction is compared with the existing methods. The performance parameters such as bit error rate, peak insertion, partial transmit sequence and selective mapping are analyzed and compared with the existing methods.

### 5.1. BER Comparison of Proposed Method with Existing Work

The BER comparison is shown in Table 2. The proposed method is compared with the available methods. the BER value for 64 carriers of SLM, PTS, SLM & PTS, SLM & Reiman, SLM & Peak windowing, clipping and filtering, and the proposed approach is 0.00452, 0.00446, 0.004321, 0.004211, 0.00423, 0.004173 respectively. It has shown that the proposed method is having minimum bit error rate than the existing methods. The bit error rate is minimized with the proposed method in comparison with the existing methods.

Table 2. Comparison of bit error rate with existing methods.

| Carrier | Selective mapping | Partial transmit sequence | SLM & PTS | SLM & Reiman | Clipping and filtering | peak windowing method | Suggested approach |
|---|---|---|---|---|---|---|---|
| 8 | $467*10^{-3}$ | $459*10^{-3}$ | $446*10^{-3}$ | $436*10^{-3}$ | $432*10^{-3}$ | $42*10^{-3}$ | $41546*10^{-3}$ |
| 12 | $461*10^{-3}$ | $452*10^{-3}$ | $442*10^{-3}$ | $432*10^{-3}$ | $423*10^{-3}$ | $419*10^{-3}$ | $4103*10^{-3}$ |
| 32 | $458*10^{-3}$ | $449*10^{-3}$ | $436*10^{-3}$ | $429*10^{-3}$ | $422*10^{-3}$ | $4186*10^{-3}$ | $4045*10^{-3}$ |
| 64 | $452*10^{-3}$ | $446*10^{-3}$ | $432*10^{-3}$ | $421*10^{-3}$ | $423*10^{-3}$ | $4173*10^{-3}$ | $3747*10^{-3}$ |

### 5.2. SNR Comparison of Proposed Method with Existing work

In comparison to existing approaches such as SLM, PTS, SLM & PTS, and SLM & Reiman, the proposed method has a maximum signal-to-noise ratio. The signal-to-noise ratio of partial transmit sequence, selective mapping & Reiman, Clipping and filtering, Peak windowing, and the proposed methods are 13.52, 13.72, 13.92, 14.42, 15.65, 17.89 and 19.69 respectively. The graphical representation of the SNR values with conventional methods is depicted as shown below in Figure 10.

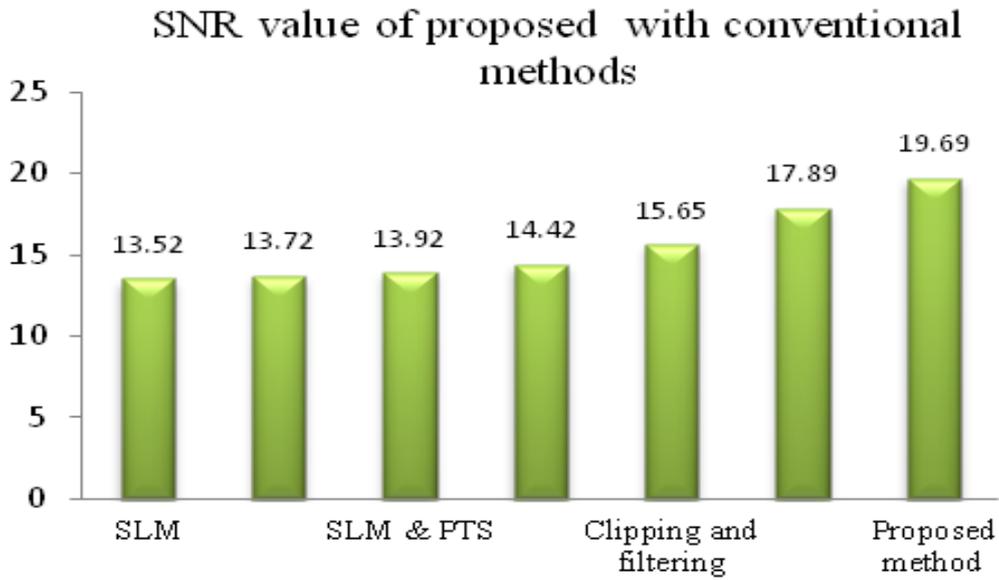

Figure 10. SNR comparison plot

### 5.3. CCDF of Proposed Method with existing work

The CCDF of SLM is approximately 7.21, whereas the CCDF of PTS, SLM & PTS, SLM-Reiman, SLM & PTS and SLM-Reiman is approximately 7.52, 6.93, 5.85, 6.93 and 5.85 respectively. The CCDF of proposed method is 4.2. It indicates that it achieves the lowest CCDF when compared to other conventional approaches. The graphical representation of the CCDF values with conventional methods is depicted as shown below in Figure 11.

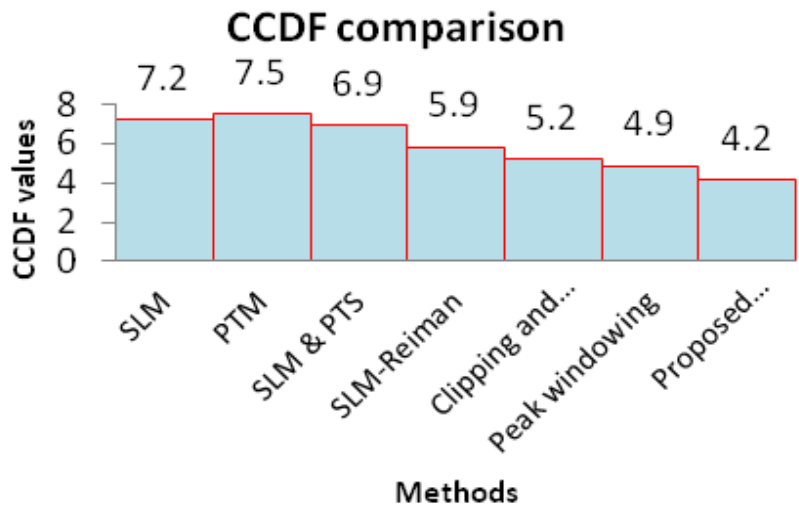

Figure 11. CCDF comparison plot

### 5.4. PAPR of Proposed Method with existing work

The PAPR of SLM, PTS, SLM & PTS, SLM-Remain and the proposed method is 6.98, 7.656, 6.77, 4.796, 4.54, 3.98 and 3.79 respectively. This shows that the PAPR value obtained with the proposed work is minimized in comparison with the existing methods. The graphical representation is shown below in Figure 12. The peak to average power ratio is minimized for the proposed method in comparison with the conventional methods. By using adaptive clipping and windowing, the bit error is reduced to 0.0037 and the PAPR is reduced to 3.7885. The proposed solution boosts the performance of wireless communication systems.

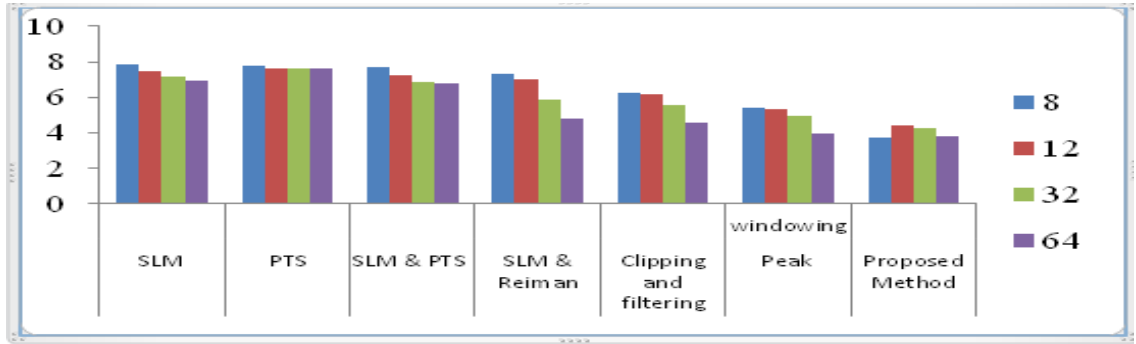

Figure 12. Comparison plot of PAPR method with existing

## 6. CONCLUSIONS

This work addresses the OFDM issues such as increased peak power, complexity in the channel, efficiency of the spectrum and robustness against the channel dispersion in the next coming generation of wireless communication systems. This research work is based on the usage of the harmonious kernel adaptive filter, Slepian-based flat-top window technique and adaptive peak window method based on harmonize clipping. This filter keeps the signal orthogonal by lowering PAPR and BER. A Slepian-based flat-top window is used to achieve noise reduction while maintaining the signal's orthogonality. This window filters out noise in the spectrum and reduces information loss. The performance parameters such as bit error rate, peak to average power ratio, and CCDF are all compared with the existing methods. The BER, SNR, CCDF and PAPR of the proposed work obtained for 64 carriers is 3747 $*10^{-3}$, 19.69, 4.2 and 3.79 $*10^{-3}$ respectively. The proposed method is having minimum BER, CCDF and PAPR than the existing techniques. The SNR value of the proposed method is higher than the conventional methods. The results obtained are analyzed and compared in this paper in sections IV and V. The proposed work is more spectral efficient and reduces the signal loss by minimizing peak to average power ratio. Hence the proposed work is a better solution for future 6G wireless communication systems.

### CONFLICTS OF INTEREST

The authors declare no conflict of interest.

### ACKNOWLEDGMENT

I would like to acknowledge JSS Academy of technical education, Bangalore, India, for providing the laboratory support to carry out the research work.

International Journal of Computer Networks & Communications (IJCNC) Vol.14, No.3, May 2022133

**AUTHORS**

Dr. Usha S.M. from Bengaluru- Karnataka, India, obtained B.E (Electronics & Communication Engineering) degree from Mysore University in the year 2000. M. Tech. in VLSI Design and Embedded Systems from VTU Belgaum in 2011 and awarded Ph.D. in Optimization and Performance Analysis of Digital Modulators from VTU Belgaum in the year 2017. She is currently working as Associate Professor at JSS Academy of Technical Education, Bengaluru. Karnataka, India. She is a member of Professional bodies such as IEEE, ISTE, and MIE. The research area of interest is VLSI design and wireless communication.

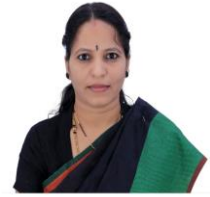

Mr. Mahesh H.B. from Bengaluru-Karnataka, India, obtained Computer Science B.E, a degree from Mysore University in the year 1996, M. Tech in Networking & Internet Engineering from VTU in the year 2004. Currently, working as an Assistant Professor at PES University, Bengaluru, and Karnataka, India. He is a member of professional bodies such as IEEE and CSE. Presently, pursuing his Ph.D. in the field of wireless networks.

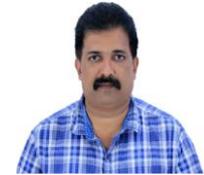